\documentclass[prl,twocolumn,aps,showpacs]{revtex4-1}

\usepackage{epsfig}
\usepackage{graphicx}
\usepackage{color}

\newcommand{\lapprox}{\stackrel{<}{\sim}}

\begin{document}

\title{Evolution of the Normal State of a Strongly Interacting Fermi Gas from a Pseudogap Phase to a Molecular Bose Gas}

\author{A. Perali $^{1}$, F. Palestini $^{1}$, P. Pieri $^{1}$, G. C. Strinati$^{1}$, J. T. Stewart $^{2}$, J. P. Gaebler $^{2}$, T. E. Drake $^{2}$, D. S. Jin$^{2}$}

\affiliation{$^{1}$Dipartimento di Fisica, Universit\`{a} di Camerino, I-62032 Camerino, Italy\\ $^{2}$JILA, NIST and University of Colorado, and Department of Physics, University of Colorado, Boulder, CO 80309-0449, USA}

\begin{abstract}
Wave-vector resolved radio frequency (rf) spectroscopy data for an ultracold trapped Fermi gas are reported for several couplings at $T_{c}$, and extensively analyzed in terms of a pairing-fluctuation theory. 
We map the evolution of a strongly interacting Fermi gas from the pseudogap phase into a fully gapped molecular Bose gas as a function of the interaction strength,  
which is marked by a rapid disappearance of a remnant Fermi surface in the single-particle dispersion.   
We also show that our theory of a pseudogap phase is consistent with a recent experimental observation as well as with Quantum Monte Carlo data of thermodynamic quantities of a unitary Fermi gas above $T_{c}$.
\end{abstract}

\maketitle

While the existence of a high-temperature superfluid phase in the BCS-BEC crossover of a strongly interacting Fermi gas is experimentally well established, important questions remain as to the nature of the gas above the superfluid transition temperature $T_{c}$.  
In particular, the question of whether or not a pseudogap state exists and how to identify it is of importance \cite{Levin-2009-review}. 
This is a question that may have relevance to the controversy surrounding the pseudogap state in the high-$T_{c}$ cuprates.  
While the origin of this state in the cuprates is a hotly debated topic, with atomic Fermi gases we can answer the simpler question of whether or not strong interactions and pairing fluctuations alone can lead to a pseudogap phase.  
This, in turn, tells us whether using such an approach to explain the pseudogap phase in the cuprates is a viable option or if other mechanisms are required.
 
As a function of increasingly strong attractive interactions, a Fermi gas exhibits a smooth crossover (called the BCS-BEC crossover), from a weakly attractive Fermi gas with a superfluid transition explained by conventional BCS theory, to a Fermi gas where interparticle attractions are so strong that the fermion pairs form molecules and the gas is well described as a molecular Bose gas with a Bose-Einstein condensation transition.  
In the BCS limit the phenomena of Cooper pairing and superfluidity occur simultaneously at the phase transition, while in the BEC limit pairing and Bose condensation are decoupled with pairing of fermionic atoms into molecules occurring well above the condensation temperature. 
The \emph{pseudogap phase} refers to the normal state of a strongly interacting Fermi gas in the center of this crossover, where it is proposed that pairs exist above the superfluid transition in analogy with the normal state of the gas in the BEC limit.  
However, unlike the pairs in the BEC limit, the pairs in the pseudogap state have many-body character with \emph{the underlying Fermi statistics playing a crucial role}, in analogy with the Cooper pairs of the BCS limit. 
A key prediction of theories of the pseudogap phase is that there should be a smooth evolution from the many-body pairs in the center of the crossover to the molecular pairs in the BEC limit  \cite{Randeria-1998,Levin-2009-review} and accordingly, in order to verify the existence of a pseudogap phase, it is critical to examine the evolution of the spectral function from the center of the crossover to the molecular limit \cite{PPSC-2002}. 
 
Based on two recent experiments, conflicting conclusions have been reached about the existence of a pseudogap state in the strongly interacting Fermi gas.  
On the one hand, thermodynamic measurements \cite{Salomon-2010} have been interpreted as well described by Fermi liquid theory, without the need for a pseudogap state.  
On the other hand, momentum-resolved rf spectroscopy \cite{Jila-Cam-2010-I}, which measures the single-particle spectral function, has been interpreted as evidence for a pseudogap state above $T_{c}$.  

In this work, we present a theoretical investigation of the pseudogap regime based on the t-matrix pairing-fluctuation approach of Ref.\cite{PPSC-2002}, addressing both the single-particle spectral function \emph{and} the thermodynamics of the gas, as a function of interaction strength in the BCS-BEC crossover.  
We find that, in the pseudogap regime, the single-particle dispersion back-bends at a wave vector $k_{L}$ near the Fermi wave vector $k_{F}$,
indicating the existence of \emph{a remnant Fermi surface} in this strongly interacting gas and the importance of Fermi statistics to the pairing.  
As interactions are increased towards the BEC limit, $k_{L}$ disappears rapidly when entering the regime of molecular pairing.  
This picture is supported by a comparison of our theoretical results, where we include the effects of the trapping potential, with new experimental data using momentum resolved rf spectroscopy to probe the gas for different interaction strengths. 
In addition, we show that the theory also reproduces the observed linear behavior in the thermodynamics.

By the experimental technique introduced in Ref.\cite{Jin-2008}, excitations of the trapped gas produced by an rf pulse are analyzed by time-of-flight imaging to determine the wave vector of the excited atoms once the trap has been switched off.
The new data are presented with an improved signal-to-noise ratio at the critical temperature $T_{c}$, which is accurately determined as the temperature where the condensate fraction disappears.
We concentrate in the coupling range $0.0 \lesssim (k_{F} a_{F})^{-1} \lesssim 1.0$, because the evolution of interest from the pseudogap state to the molecular Bose gas occurs on the positive side of the resonance.
Here, $a_{F}$ is the scattering length  associated with  the Fano-Feshbach resonance and $k_{F}$ is given by $\hbar^{2} k_{F}^{2}/(2m) = E_{F} = \hbar \omega_{0} (3N)^{1/3}$, where $\hbar$ is Planck constant, $m$ the atom mass, $N$ the total number of atoms, and $\omega_{0}$ the average trap frequency (we set $\hbar = 1$).

Ultracold Fermi gases are peculiar systems, in that their interparticle coupling can be increased to the point when a description in terms of a gas of molecular bosons holds, for which a real gap exists in the single-particle spectra. This molecular (two-body) physics is of no interest in the context of the pseudogap, in a similar fashion of molecular binding in vacuum being distinct from Cooper pairing at finite density in the presence of a Fermi surface
(cf. footnote 18 of Ref.\cite{BCS-1957}). 
The question then arises about what fermionic feature distinguishes the pseudogap from the molecular phase.
We shall find that the back-bending of the dispersion curves obtained from the single-particle spectral function $A(k,\omega)$ (with wave vector $k$ and frequency $\omega$) occurs at a wave vector $k_{L}$ which remains close to $k_{F}$ over a wide coupling range even when approaching the molecular limit.
We refer to this special wave vector as $k_{L}$ because it is reminiscent of the Luttinger theorem \cite{Luttinger-1960}, according to which in a normal Fermi liquid the radius $k_{F}$ of the Fermi sphere is unaffected by the interaction.

\begin{figure}[t]
\begin{center}
\includegraphics[angle=0,width=8.5cm]{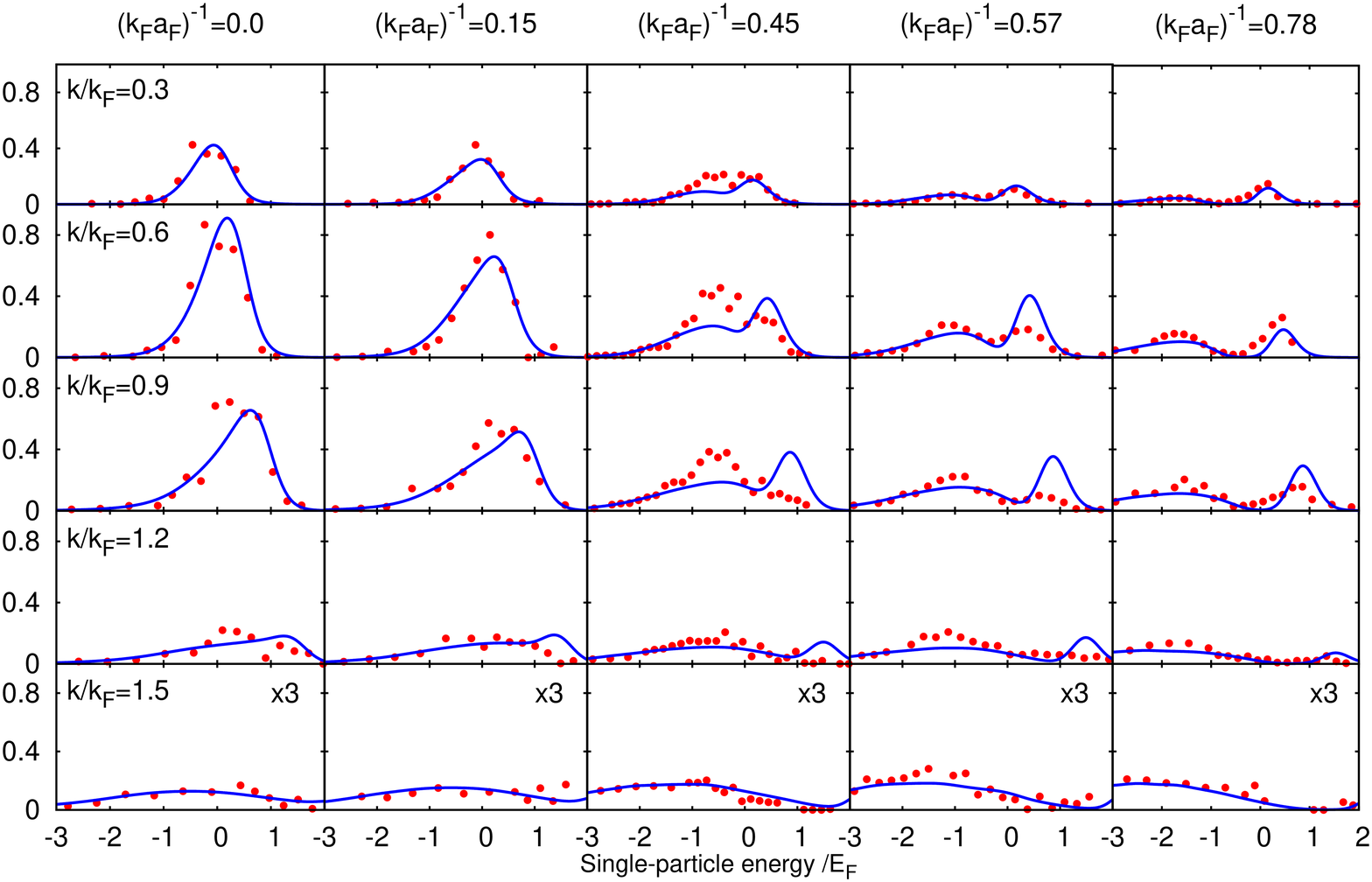}
\caption{ Experimental (circles) and theoretical (full lines) EDC for the trap at $T_{c}$, for several couplings and wave vectors.}
\label{fig1}
\end{center}
\end{figure} 

Figure~\ref{fig1} compares the experimental and theoretical energy distribution curves (EDC) at $T_{c}$ for five different couplings in the window of interest
(see Ref.\cite{SOM} for details). 
We emphasize that the experimental data bear on \emph{an absolute normalization}, in that only the integral over wave vector and energy of the EDC curves (and not the separate spectra) has been normalized to unity \cite{SOM}.
For this reason, there is no independent normalization in the various panels at different $k$.
This renders quite stringent the comparison with the corresponding theoretical calculations, which in turn contain no adjustable parameters.
Good agreement results from this comparison. 
In particular, the theoretical calculations well reproduce the asymmetry of the experimental curves between positive and negative energies, in addition to the peak positions, widths and heights (note how the latter change by about one order of magnitude from small to large $k$). Note further the excellent agreement between the theoretical and experimental negative energy tails, and the gradual flattening of the EDC curves for increasing coupling due to the increase of intrapair correlations.

\begin{figure}[t]
\begin{center}
\includegraphics[angle=0,width=8.5cm]{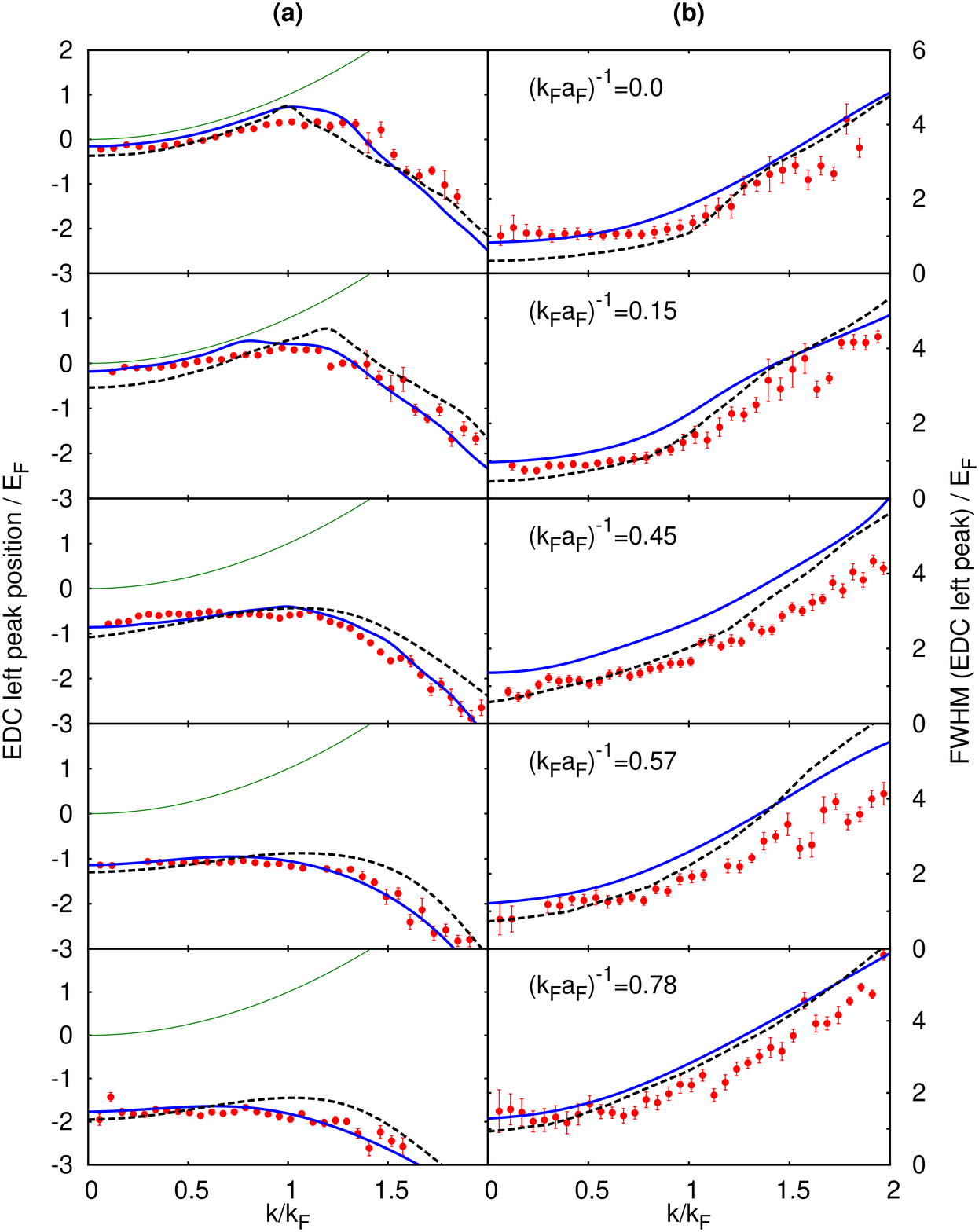}
\caption{(a) Dispersions and (b) widths of the low-energy EDC peak. Experimental data (circles) and theoretical calculations for the trap (full lines) are shown for 
               the same couplings of Fig.~\ref{fig1}, and compared with the contribution from the radial shell with the largest particle number (dashed lines). 
               In the left panels the free-particle dispersion $k^{2}/(2m)$ is also reported for comparison (thin full lines).}
\label{fig2}
\end{center}
\end{figure} 

In Fig.~\ref{fig2} the dispersion and full width at half maximum of the peak at lower energies are reported over a dense set of $k$ values 
for the same couplings of Fig.~\ref{fig1}, and compared with our theoretical calculations. 
Note that a characteristic \emph{back-bending} is revealed from these dispersions \cite{Schneider-2010}.
This kind of back-bending is typical of a BCS-like dispersion, and is associated with the presence of a pseudogap in a
strongly interacting Fermi system \cite{Jila-Cam-2010-I,PPSC-2002,Levin-2009,Bulgac-2009,Ohashi-2009}.
In addition, the large values of the widths (which are at least of the order of $E_{F}$) and their asymmetric behavior 
between $k < k_{F}$ and $k > k_{F}$ are associated with strong deviations from the expected behavior of a normal Fermi liquid (which requires instead the quasi-particle widths to be vanishingly small at $k_{F}$ \cite{Nozieres-1964}), 
and confirm the fact that single-particle states in this region constitute poor quasi-particles.
Large values of the widths are not surprising in the context of the pseudogap physics that results from pairing fluctuations \cite{PPSC-2002}. 
Large widths were also obtained by the self-consistent t-matrix approach of Ref.~\cite{Zwerger-2009}, which however masked 
the occurrence of a pseudogap near $k_F$.

It is relevant to discuss how trap averaging affects the above results, because different radial shells in the trap correspond to different locations in 
the coupling-vs-temperature phase diagram of the homogeneous system.
A reasonable hypothesis is that the radial shell with the largest particle number (whose radius $r_{\mathrm{max}}$ is estimated to be $(0.5 - 0.6) R_{F}$ where
$R_{F}=[2E_{F}/(m\omega_{0}^{2})]^{1/2}$ is the Thomas-Fermi radius) contributes most to the total signal. 
The dispersions and widths contributed by this shell at $r_{\mathrm{max}}$ are represented by dashed lines in Fig.~\ref{fig2}, which show good agreement with the complete calculation. 
This indicates that both the back-bending of the dispersion relations and the associated large widths are not an artifact of trap averaging.

\begin{figure}[t]
\begin{center}
\includegraphics[angle=0,width=8.8cm]{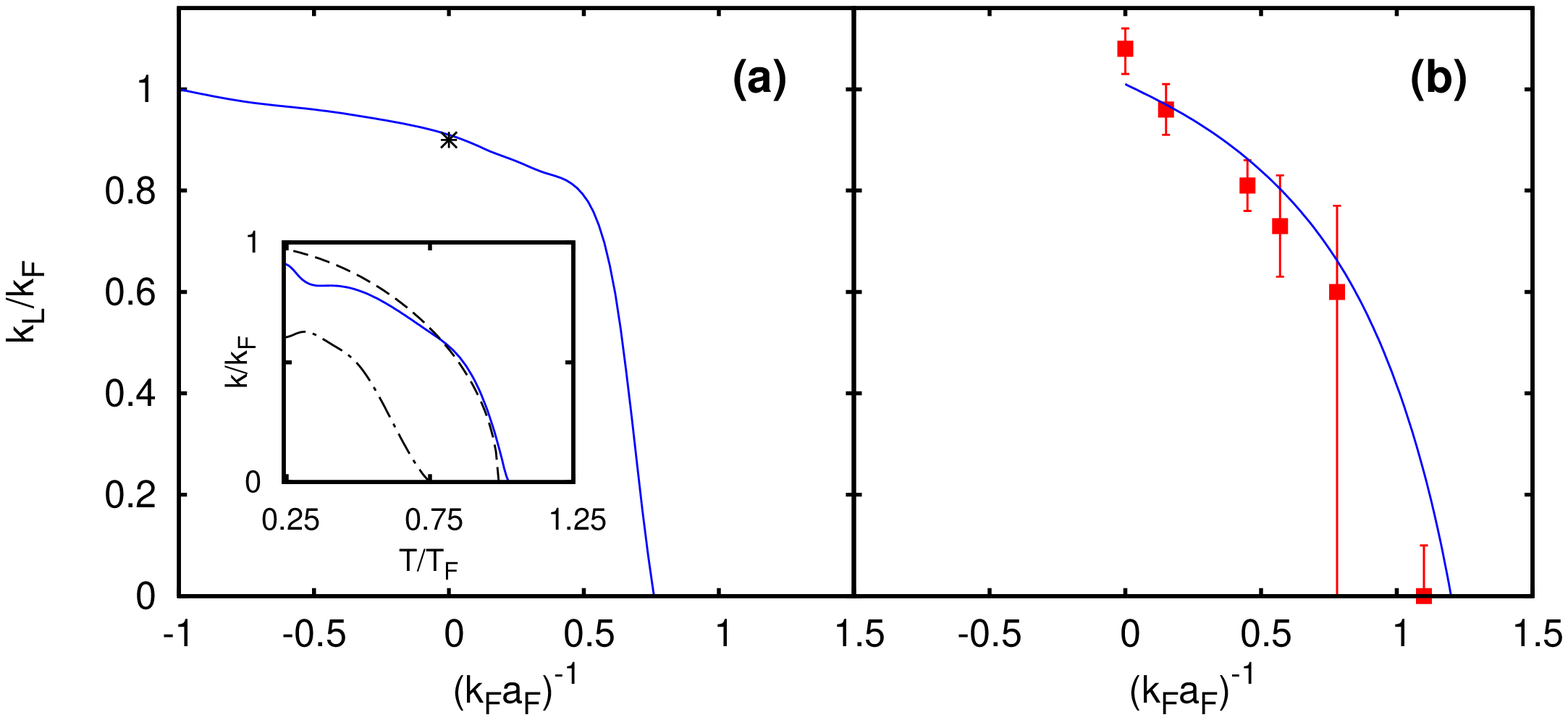}
\caption{(a) Coupling dependence of the Luttinger wave vector $k_L$ for a homogeneous system at $T_{c}$, according to the theory 
              of Ref.\cite{PPSC-2002} (full line)
              [the value at unitarity from the QMC calculation of Ref.\cite{Carlson-2005} is also reported (star)].
              The inset shows the temperature dependence of $k_L$ at unitarity (full line), and compares it with those obtained from the temperature 
              dependence of the chemical potential of the non-interacting (dashed line) and interacting (dashed-dotted line) systems.
              (b) Theoretical (full line) and experimental (squares) coupling dependence of $k_L$ for the trap system at $T_{c}$.}
\label{fig3}
\end{center} 
\end{figure}

Despite  these deviations from the behavior of a normal Fermi liquid, in the experimental data and theoretical calculations 
there yet appears a feature which is preserved from the physics of a Fermi liquid.
That is the Luttinger wave vector $k_{L}$ where the back-bending occurs, which is plotted at $T_{c}$ vs $(k_{F} a_{F})^{-1}$ in Fig.~\ref{fig3}, for a homogeneous [panel (a)]
and trapped [panel (b)] system.

Figure~\ref{fig3}(a) shows that for a homogeneous system $k_{L}$ drops rapidly to zero when $(k_{F} a_{F})^{-1} \simeq 0.75$, where the pseudogap in $A(k,\omega)$ turns into a real gap and the molecular limit is reached.
Accordingly, we identify the boundary between the pseudogap and molecular phases where this drop occurs.
 Along this evolution into the molecular regime, the disappearance of the underlying Fermi surface about occurs when the molecular size becomes smaller than the interparticle spacing. 
The existence of a remnant Fermi surface with an enclosed volume consistent with Luttinger theorem was already pointed out by ARPES experiments for the pseudogap phase of high-$T_{c}$ superconductors \cite{Shen-1998}, but its importance for delimiting the pseudogap region was not appreciated in that context 
\cite{KM-2000} because the interparticle interaction could not be controlled.
The inset of Fig.~\ref{fig3}(a) shows the temperature dependence of $k_{L}$ calculated for a homogeneous system at unitarity (full line).
At high temperatures when the pseudogap closes up, we have identified $k_{L}$ as the value where the dispersion of the peak at lower energy in 
$A(k,\omega)$ crosses the chemical potential \cite{SOM}. 
This does not contradict our argument that at low temperatures the presence of a pseudogap requires an underlying Fermi surface, since at high temperatures the underlying Fermi surface of a Fermi liquid is not related to a pseudogap.
The plot also shows the temperature dependence of $k_{\mu^{0}} = \sqrt{2 m \mu^{0}(T)}$ (dashed line) and $k_{\mu} = \sqrt{2 m \mu(T)}$ (dashed-dotted line), where $\mu^{0}(T)$ and $\mu(T)$ are the chemical potentials of the non-interacting and interacting Fermi systems, in the order, at the temperature $T$.
Note that $k_{L}$ about coincides with $k_{\mu^{0}}$, while  $k_{\mu}$ is not related with $k_{L}$.

Figure~\ref{fig3}(b) shows the coupling dependence of $k_{L}$ at $T_{c}$ for the trapped system, for which the theoretical predictions can be directly compared with the experimental data (the latter are obtained by a BCS-like fit to the dispersions of Fig.~\ref{fig2}(a), as explained in Ref.\cite{SOM}).
The good comparison that results between theory and experiment confirms our identification of $k_{L}$ as the relevant quantity for identifying the remnant Fermi characteristics 
of the system in the pseudogap phase. 

\begin{figure}[t]
\begin{center}
\includegraphics[angle=0,width=5.0cm]{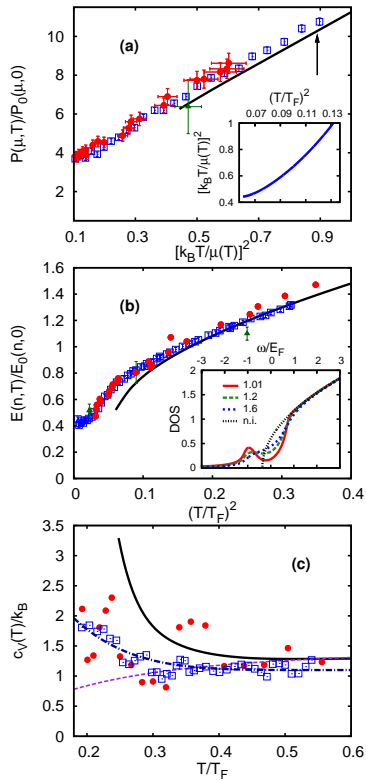}
\caption{Thermodynamics of a homogeneous Fermi gas at unitarity. 
             (a) Pressure vs $[k_{B} T / \mu(T)]^{2}$: Experimental data from Ref.\cite{Salomon-2010} (circles) are compared with QMC data from Refs.\cite{Bulgac-2006}
             (squares) and \cite{Burovski-2006} (triangles), and with the t-matrix (full line). In the inset, the variable $[k_{B} T / \mu(T)]^{2}$ is transformed to $(T/T_{F})^{2}$         
             according to the t-matrix.
             (b) Energy vs $(T/T_{F})^{2}$ at fixed density: Experimental data from Ref.\cite{Salomon-2010-II} (circles) are compared with QMC data from Refs.\cite{Bulgac-2006}
             (squares) and \cite{Burovski-2006} (triangles), and with the t-matrix (full line). The inset shows the density of states per spin (in units of $m k_{F}/(2 \pi)^{2}$) for several temperatures 
             in units of $T_{c}$ according to the t-matrix, and contrasts it with the non-interacting (n.i.) result.
             (c) Specific heat per particle vs $T/T_{F}$ obtained from the t-matrix (full line), the experimental data of Ref.\cite{Salomon-2010-II} (circles), and the QMC data of Ref.\cite{Bulgac-2006}
             (squares) -
             the dotted line is a guide to the eye for the QMC data. The behavior of the non-interacting Fermi gas (broken line) is reported for reference \cite{SOM}.}
\label{fig4}
\end{center}
\end{figure}

However, the occurrence of a pseudogap for a unitary Fermi gas above $T_{c}$ has recently been questioned, following a result reported in Ref.\cite{Salomon-2010} where a linear dependence of the equation of state as a function of $[k_{B} T / \mu(T)]^{2}$ ($k_{B}$ being Boltzmann constant) was fitted by the Fermi-liquid equation of state and then interpreted \cite{Shin-2010} as evidence that the Fermi-liquid theory with no pseudogap can describe a unitary Fermi gas above $T_{c}$.
To compare with the data of Ref.\cite{Salomon-2010} and resolve this controversy, we have used the theoretical approach of Ref.\cite{PPSC-2002}, which contains a robust pseudogap associated with a non-Fermi-liquid behavior consistent with the data obtained by momentum resolved rf spectroscopy, also to calculate the thermodynamic properties of a homogeneous system above 
$T_{c}$. 
Figure~\ref{fig4}(a) reports the pressure in the grand-canonical ensemble vs $[k_{B} T / \mu(T)]^{2}$ as in Ref.\cite{Salomon-2010}, and shows that the \emph{linear behavior} seen in the experimental data and QMC calculations also results from our t-matrix approach, both above and below the temperature at which the pseudogap appears (indicated by the vertical arrow).
The inset of Fig.~\ref{fig4}(a) shows that this linear behavior can be ascribed to the pronounced temperature dependence of the chemical potential, because a non-linear behavior results when transforming $[k_{B} T / \mu(T)]^{2}$ to $(T/T_{F})^{2}$ over the relevant range.
The same change of variables can be performed in the experimental \cite{Salomon-2010-II} and QMC \cite{Bulgac-2006,Burovski-2006} data, to obtain the total energy in the canonical ensemble as a function of $(T/T_{F})^{2}$ reported in Fig.~\ref{fig4}(b).
This shows that in the new variable the linear behavior is lost.
 
Yet, it remains difficult to appreciate directly from this thermodynamic quantity the presence of a pseudogap in a unitary Fermi gas above $T_{c}$ even by the t-matrix calculation, despite the fact that a pseudogap is clearly present in the single-particle density of states obtained by the t-matrix as shown in the inset of Fig.~\ref{fig4}(b) where deviations from the non-interacting behavior
$\sqrt{(\omega + \mu(T_{c})) / E_{F}}$ are evident.
Accordingly, by suitable numerical differentiation of the energy data we have obtained in Fig.~\ref{fig4}(c) the \emph{specific heat} vs $T/T_{F}$.
A sharp upturn of this thermodynamic quantity, beginning at a temperature $T^{*}$ well above $T_{c}$ where the pseudogap sets in, results clearly from the t-matrix calculation, and it is also visible from the QMC data at the corresponding value of $T_{c}$.

The experimental data in Fig.~\ref{fig4}(c) appear too scattered to draw definite conclusions about the presence of the upturn and thus of a pseudogap above $T_{c}$.
It should be mentioned, however, that a similar upturn of the specific heat at a temperature $T^{*}$ above $T_{c}$ was measured in underdoped high-$T_{c}$ cuprates and interpreted as revealing the onset of the pseudogap regime, whereby a ``residual superconductivity'' remains far above $T_{c}$ \cite{SH-HTcSC-2009}.    

 In conclusion, we have provided clear experimental and theoretical evidence for non-Fermi-liquid behavior in the normal phase of a strongly interacting Fermi gas, which we have qualified in terms of a pseudogap picture.
We have further shown that this picture, that appears evident in the single-particle dynamics, is also consistent with the
thermodynamic behavior of the system.       

\acknowledgments
We acknowledge financial support from the NSF and from the Italian MIUR under contract PRIN-2007 ``Ultracold Atoms and Novel Quantum Phases''. 



\newpage

{\bf Supplemental material: ``Evolution of the Normal State of a Strongly Interacting Fermi Gas from a Pseudogap Phase to a Molecular Bose Gas''}
\\
\\
{\em We provide details of the theoretical calculations of the wave-vector resolved rf signal and a description of the experimental procedures. We also add information about the theoretical analysis of the experimental data.}


\begin{center}
{\bf Pairing-fluctuation theory}
\end{center}
\vspace{-0.2cm}

\noindent
The theoretical approach of Ref.~\cite{S-PPSC-2002} is based on a diagrammatic t-matrix approximation, whereby the fermionic single-particle self-energy $\Sigma(\mathbf{k},\omega)$ includes pairing fluctuations.
We have used that approach here to calculate the single-particle spectral function $A(\mathbf{k},\omega)$ for the homogeneous case:
\begin{equation}
A(\mathbf{k},\omega) \, = \, - \frac{1}{\pi} \, \frac{\mathrm{Im}\Sigma(\mathbf{k},\omega)}
{[\omega - \xi_{\mathbf{k}} - \mathrm{Re}\Sigma(\mathbf{k},\omega)]^{2} \, + \, 
[\mathrm{Im}\Sigma(\mathbf{k},\omega)]^{2}}                                                              \label{single-particle-spectral-function}
\end{equation}

\noindent
where $\xi_{\mathbf{k}} = \mathbf{k}^{2}/(2m) - \mu$.
For given wave vector $\mathbf{k}$, the frequency structure of the real and imaginary parts of $\Sigma(\mathbf{k},\omega)$ determines the positions and widths of the peaks in $A(\mathbf{k},\omega)$, and is thus responsible for the nontrivial shape of the dispersions of these peaks vs $k = |\mathbf{k}|$. 

\begin{center}
{\bf Wave-vector resolved rf spectroscopy}
\end{center}
\vspace{-0.2cm}

\noindent
When final-state effects in the rf transition \cite{S-PPS-2008,S-PPS-2009} can be neglected (like for the case of $^{40}\mathrm{K}$ used in the experiment), 
the rf signal in the normal phase is given by \cite{S-PPS-2009}:
\begin{equation}
\mathrm{RF}(\tilde{\omega})=\frac{1}{\pi N}\int \!\! d \mathbf{r} \int \!\! \frac{d \mathbf{k}}{(2 \pi)^{3}} A(\mathbf{k}, \xi(\mathbf{k};\mathbf{r}) -  \tilde{\omega})f(\xi(\mathbf{k};\mathbf{r}) -\tilde{\omega}).             
\label{total-RF-signal}
\end{equation}

\noindent
Here, $\tilde{\omega} = \omega_{\mathrm{rf}} - \omega_{a}$ is the detuning frequency where $\omega_{\mathrm{rf}}$ is the frequency of the rf photon and $\omega_{a}$ the atomic hyperfine frequency, $\mathbf{r}$ the position in the trap, $\xi(\mathbf{k};\mathbf{r}) = \mathbf{k}^{2}/(2m) - \mu + V(\mathbf{r})$ a local energy with trapping potential
$V(\mathbf{r}) = m \left( \omega_{x} x^{2} + \omega_{y} y^{2} + \omega_{z} z^{2} \right)/2$, and $f(\epsilon) = \left( e^{\epsilon/(k_{B}T)} + 1 \right)^{-1}$ the Fermi function.
[The prefactor in Eq.(\ref{total-RF-signal}) is chosen to make the total area of the rf signal equal unity.]
Equation (\ref{total-RF-signal}) is based on a local-density approximation where contributions of adjacent shells in the trap are separately considered.

The rf signal can be analyzed into its individual $\mathbf{k}$-components to compare with the experimental technique of Ref.~\cite{S-Jin-2008}.
The resulting \emph{wave-vector resolved rf signal\/} is obtained by dropping the $\mathbf{k}$-integration and considering one $\mathbf{k}$-component at a time.
More precisely, the selection is made over the magnitude $k = |\mathbf{k}|$ while $\hat{k} = \mathbf{k}/k$ is integrated over the solid angle, yielding:
\begin{equation}
\mathrm{RF}(k,\tilde{\omega})=\frac{48 k^{2}}{\pi^{2}}\int_{0}^{\infty} \!\!\!
dr r^{2}A(k, \xi(k;r) - \tilde{\omega})f(\xi(k;r) - \tilde{\omega})        
\label{partial-RF-signal}
\end{equation}
where the factor $k^{2}$ is from the spherical integration and $r = |\mathbf{r}|$ is the radial position in the trap.
The prefactor here results by expressing wave vectors in units of $k_{F}$, energies in units of $E_{F}$, and radial positions in units of the Thomas-Fermi radius 
$R_{F}=[2E_{F}/(m\omega_{0}^{2})]^{1/2}$ where $\omega_{0}=(\omega_{x}\omega_{y}\omega_{z})^{1/3}$ is the average trap frequency.

Finally, to obtain an expression that can be directly compared with the experimental EDC spectra, it is sufficient to express the frequency $\tilde{\omega}$ in Eq.~(\ref{partial-RF-signal}) in terms of the single-particle energy $E_{s} = k^{2}/(2m) - \tilde{\omega}$ via the relation $\xi(k;r) -  \tilde{\omega} = E_{s} - \mu(r)$, where $\mu(r) = \mu - m \omega_{0}^{2}r^{2}/2$ is the local chemical potential in the harmonic trap.
This yields eventually:
\begin{equation}
\mathrm{EDC}(k,E_{s})= \frac{48k^{2}}{\pi^{2}}\!\!\int_{0}^{\infty} \!\!\!\! dr \, r^{2} A(k,E_{s} - \mu(r)) f(E_{s} - \mu(r)).          \label{theoretical-ECC-curves}
\end{equation}

\noindent
The numerical results obtained from Eq.(\ref{theoretical-ECC-curves}) are then convoluted by a Gaussian broadening with a rms of about $0.25 E_{F}$, corresponding to the experimental resolution.

When the interparticle interaction is switched off, $A(k, E_{s} - \mu(r);r)$ is given by $\delta(E_{s} - k^{2}/(2m))$.
This defines the \emph{zero\/} of the single-particle energy in the EDC curves as the energy of an isolated atom at rest.
The chemical potential has thus disappeared from the free-particle branch $k^{2}/(2m)$, which remains positive for all $k$ and can be used to reckon the value of the pseudogap.

\begin{center}
{\bf Experimental procedures} 
\end{center}
\vspace{-0.2cm}

\noindent
We refer to Ref.~\cite{S-Jila-Cam-2010-I} for a detailed description of the experimental techniques and procedures. Here, we add a few comments that are specifically relevant to the data presented in the main paper.

These data are taken at $T/T_{c}= 1.0 \pm 0.1$, where $T_{c}$ is determined in the trapped system by the vanishing of the measured condensate fraction.
Note, however, that, because the density of the trapped gas is spatially inhomogeneous, the local critical temperature decreases away from the cloud center. 

\begin{figure}[t]
\begin{center}
\includegraphics[angle=0,width=6.5cm]{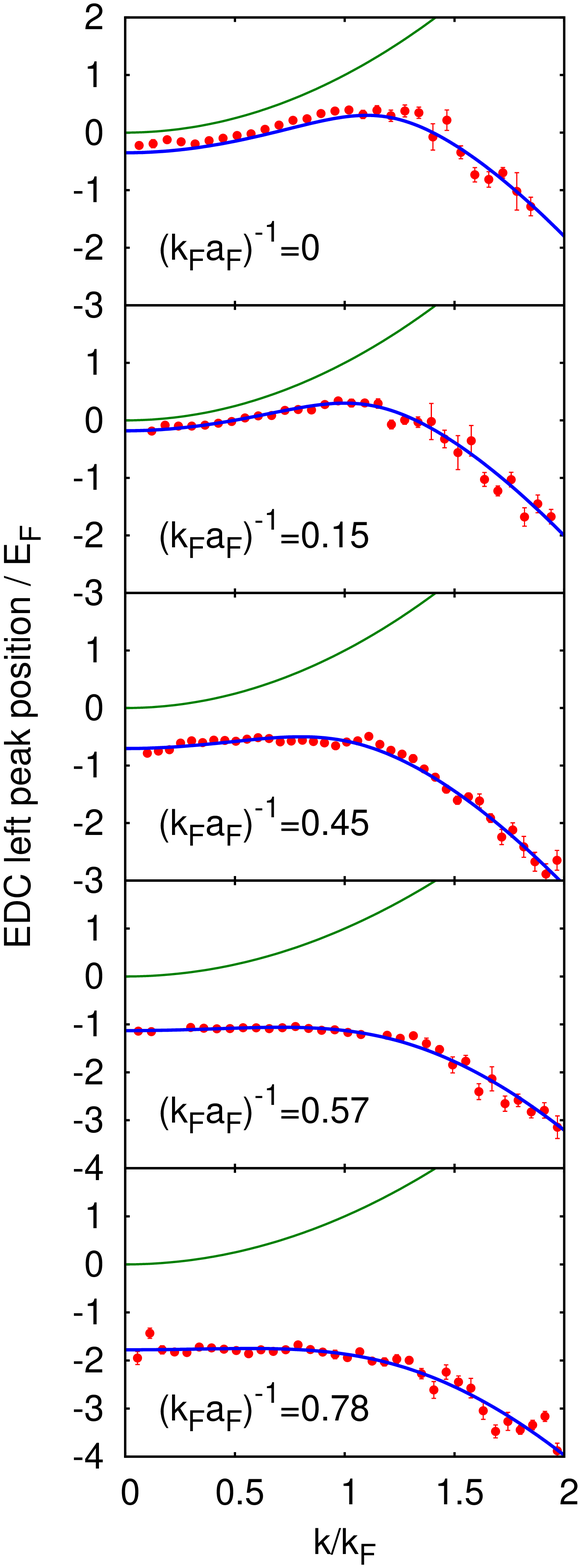}
\caption{$\chi^{\;\;2}-$fit to the experimental data. The experimental data of Fig.~2(a) of the main paper (circles) are fitted over the interval $0.0 \le k/k_{F} \le 2.0$ by the BCS-like  
              dispersion of Eq.(\ref{BCS-like-dispersion}) (full lines). The free-particle dispersion $k^{2}/(2m)$ is also shown for comparison (thin full lines).}
\label{figS1}
\end{center}
\end{figure}

It was already remarked in the main paper that an \emph{absolute\/} comparison can be made between the experimental and theoretical EDC curves at given coupling and wave vector, in such a way that only their overall integral over $k$ and $E_{s}$ (and not the individual EDC curves) are normalized to unity.
This is possible because the experimental radio frequency data (from which the EDC curves are obtained) were taken for the first time over a wide range of $\tilde{\omega}$, 
such that their long high-frequency tail could be determined and accurately compared with the $\tilde{\omega}^{-3/2}$ behavior predicted theoretically \cite{S-PPS-2008}.

The experimental data reported in Fig.~2(a) of the main paper have been analyzed in terms of the BCS-like dispersion:
\begin{equation}
E_{s}(k) \, = \, \tilde{\mu} \, - \, \sqrt{ \left( \frac{k^{2}}{2m} \, - \, \frac{k_{L}^{2}}{2m} \right)^{2} \, + \, \tilde{\Delta}^{2} }       \label{BCS-like-dispersion}
\end{equation}
where $k_{L}\approx k_{F}$ is the special wave vector about which the back-bending occurs and $\tilde{\mu}$ accounts for an overall (upward) displacement of the dispersion curves.
Note that, contrary to the homogeneous case, in a trap $\tilde{\mu}$ is \emph{not} related to the value of the thermodynamic chemical potential close to $T_{c}$.

A $\chi^{2}-$analysis of the data in the interval $0.0 \le k/k_{F} \le 2.0$ yields the fits shown in Fig.~\ref{figS1}. 
The five values of $k_{L}$ thus obtained have been reported (together with the corresponding error bars) in Fig.~3(b) of the main paper (there, the additional value for the coupling
$(k_{F} a_{F})^{-1} = 1.1$ has been inferred from the experimental data reported in Ref.\cite{S-Jin-2008}).

The same analysis also shows that pairs of $(\tilde{\mu},\tilde{\Delta})$ with $\tilde{\mu} - \tilde{\Delta} = \mathrm{constant}$ produce comparable $\chi^{2}$ tests.
From Eq.(\ref{BCS-like-dispersion}) we note that $\tilde{\mu} - \tilde{\Delta} = E_{s}(k_{L}) \equiv E_{\mathrm{max}}$ corresponds to the maximum value of $E_{s}(k)$.
For the five couplings here considered we obtain the values $E_{\mathrm{max}}/E_{F} = (0.40,0.24,-0.5,-1.1,-1.8)$, in the order.

Using these values, one can extract a rough estimate of a (trap averaged) pseudogap energy, by relating them with the free-particle dispersion $k^{2}/(2m)$ at $k_{L}$.
The $k^{2}$-dispersion can, in fact, be considered as \emph{a lower bound\/} to the dispersion of the upper branch in the EDC curves, which results from the two-peak structure of $A(k,\omega)$ in the presence of a pseudogap \cite{S-PPSC-2002} and behaves like $k^{2}/(2m)$ for $k_{F} \ll k$. 
[In the analysis of the experimental data the visibility of this upper branch is suppressed by the presence of the Fermi function.]
The values we obtain for $[k_{L}^{2}/(2m) - E_{\mathrm{max}}]/2$ are $(0.38,0.34,0.58,0.82,1.08) E_{F}$ for the five couplings of Fig.~\ref{figS1}, which are in line with the expected trend for the
pseudogap of a homogeneous system 
(cf. Fig.~17 of Ref.\cite{S-PPSC-2002}).
 
\begin{center}
{\bf Determination of $k_{L}$ for a homogeneous system}
\end{center}
\vspace{-0.2cm}

A comment is in order about the procedure for identifying the Luttinger wave vector $k_{L}$ for a homogeneous system, as reported in Fig.~3(a) of the main paper.

Quite generally, $A(\mathbf{k},\omega)$ given by Eq.(\ref{single-particle-spectral-function}) has a pronounced peak when the following condition is satisfied
\begin{equation}
\omega - \xi_{\mathbf{k}} - \mathrm{Re}\Sigma(\mathbf{k},\omega) \, = \, 0             \label{peak-condition}
\end{equation}

\noindent
and provided $\mathrm{Im}\Sigma(\mathbf{k},\omega)$ is sufficiently small.
In a BCS-like situation we write:
\begin{equation}
\Sigma(\mathbf{k},\omega) \, \approx \, \frac{\Delta_{\mathrm{pg}}^{2}}{\omega + i \eta + \xi_{\mathbf{k}} + \delta\mu} \, + \, \delta\mu              \label{BCS-like-self-energy}
\end{equation}

\noindent
with $\eta = 0^{+}$ and $\delta\mu = \mu - \mu_{L}$ where $\mu_{L} = k_{L}^{2}/(2m)$.
Note that the shift $\delta\mu$ is the part of the self-energy which is responsible for the persistence of a remnant Fermi surface about the (temperature dependent) Fermi wave vector of the 
underlying non-interacting system.
A combination of Eqs.(\ref{peak-condition}) and (\ref{BCS-like-self-energy}) then yields:
\begin{equation}
\omega \approx - \sqrt{\left( \xi_{\mathbf{k}} + \delta\mu \right)^{2} + \Delta_{\mathrm{pg}}^{2}} = 
                          - \sqrt{\left( \frac{k^{2}}{2m} - \frac{k_{L}^{2}}{2m} \right)^{2} + \Delta_{\mathrm{pg}}^{2}}
\end{equation}

\noindent
for the lower branch (which has the largest spectral weight for $k \lapprox k_{L}$), where $\omega$ is measured with respect to the chemical potential.
The maximum value $\omega \approx -  \Delta_{\mathrm{pg}}$ is for $k = k_{L}$ where the back-bending occurs.
For increasing temperature such that $ \Delta_{\mathrm{pg}}$ closes up eventually, Eq.(\ref{peak-condition}) yields accordingly $k = k_{L}$ for $\omega = 0$, which corresponds to the familiar condition for a Fermi liquid \cite{S-LP-Stat_Phys}.

\begin{figure}[t]
\begin{center}
\includegraphics[angle=0,width=7.0cm]{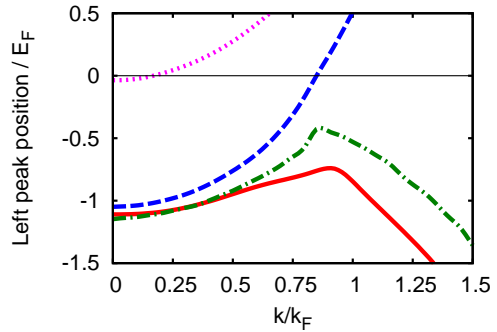}
\caption{Evolution in temperature of the dispersion $\omega(k)$ obtained by following the peak at lower energy in $A(k,\omega)$ for a homogeneous system at unitarity.
              The four curves correspond to temperatures $T/T_{c} = (1.0,1.2,1.65,4.0)$ from bottom to top.}
\label{figS2}
\end{center}
\end{figure}

Figure \ref{figS2} shows a typical temperature evolution of the dispersion $\omega(\mathbf{k})$ obtained by following the peak at lower energy in $A(\mathbf{k},\omega)$, from which the value of $k = k_{L}$ is extracted according to the above criterion.
At sufficiently high temperatures when the dispersion $\omega(\mathbf{k})$ crosses zero, $k_{L}$ is seen to quickly converge to the value associated with the temperature dependent chemical potential of the non-interacting Fermi system.

\begin{center}
{\bf Additional theoretical analysis of the experimental data}
\end{center}
\vspace{-0.2cm}

\begin{figure}[t]
\begin{center}
\includegraphics[angle=0,width=7.0cm]{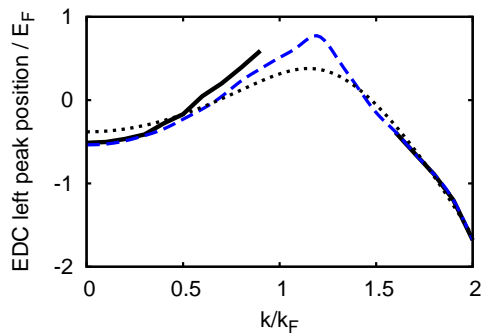}
\caption{The dispersion (two arcs drawn by full lines) of the low-frequency peak of $A(k,\omega)$ for a homogeneous system with the density of the shell at $r=r_{\mathrm{max}}$ 
              for the coupling $(k_Fa_F)^{-1}=0.15$, is compared with the corresponding dispersion (dashed line) obtained multiplying $A(k,\omega)$ by $f(\omega)$. 
              A BCS-like fit to the two arcs is also shown (dotted line).} 
\label{figS3}
\end{center}
\end{figure}

In the expression (\ref{theoretical-ECC-curves}) the presence of the Fermi function $f(\omega)$ may be of considerable help for the analysis of the dispersion of the low-$\omega$ peak, in situations when two broad non-Lorentzian structures in $A(k,\omega)$ merge together over a limited range of $k$.
This is because multiplication of $A(k,\omega)$ by $f(\omega)$ in that expression acts effectively as a ``filter'' for the low-$\omega$ structures of $A(k,\omega)$, in particular for those values of $\omega$ through which the back-bending occurs in the dispersion.

\begin{figure}[t]
\begin{center}
\includegraphics[angle=0,width=7.0cm]{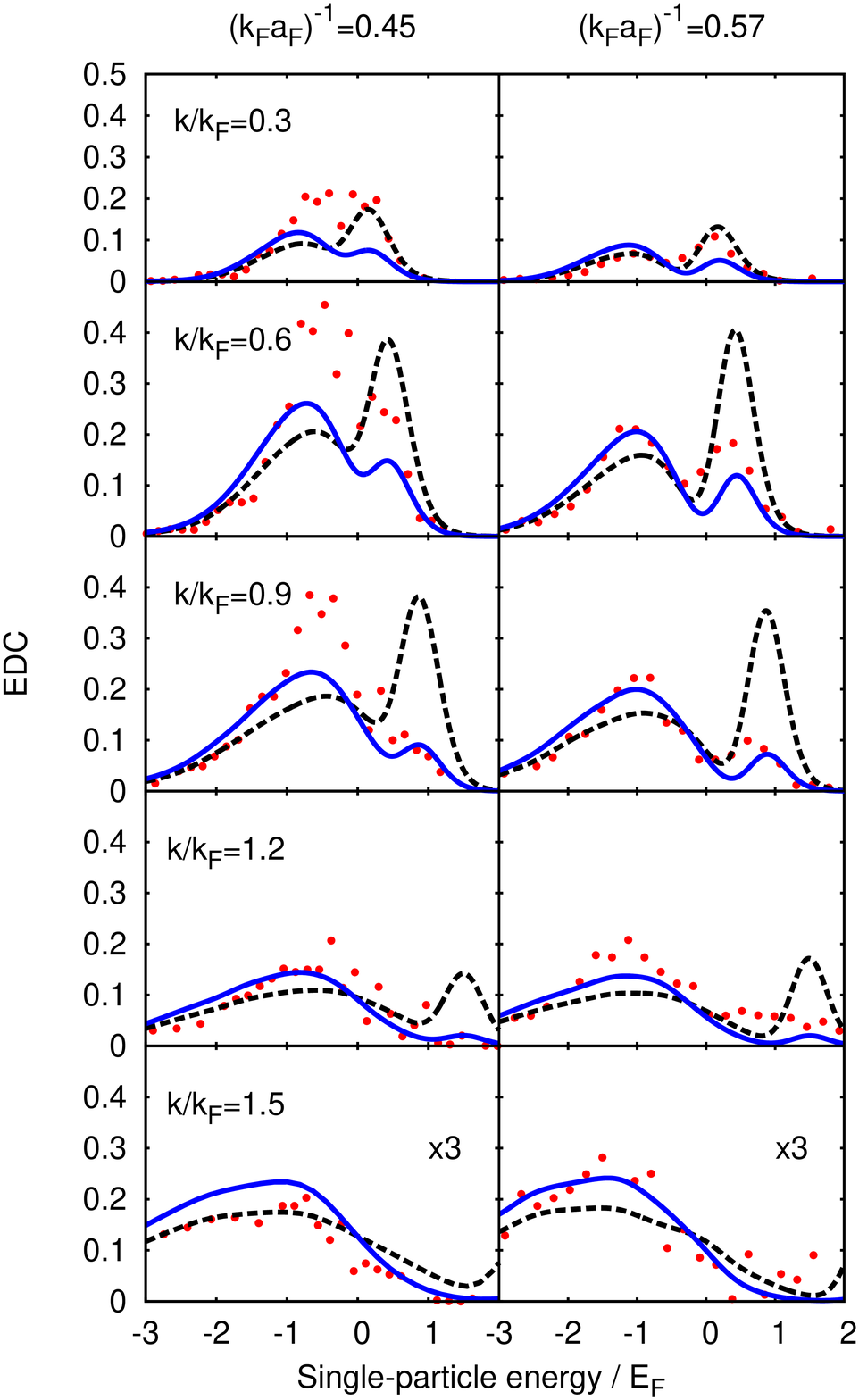}
\caption{The experimental EDC (circles) for the two couplings 0.45 and 0.57 are reproduced from Fig.~1 of the main paper, and compared with theoretical calculations (full lines)
              in which the temperature in the Fermi function has been decreased to $0.7 T_{c}$.
              The theoretical curves reported in Fig.~1 of the main paper are also reproduced here (dashed lines).} 
\label{figS4}
\end{center}
\end{figure}

This is shown explicitly in Fig.~\ref{figS3}, where the dispersion of the low-$\omega$ peak of $A(k,\omega)$ (corresponding to a homogeneous system with the density of the shell at $r_{\mathrm{max}}$ for the trap coupling $(k_Fa_F)^{-1}=0.15$) is drawn (full line) only for those values of $k$ for which two peaks in $A(k,\omega)$ appear clearly distinguishable.
This procedure results in two arcs separated by an empty window.
A single BCS-like fit (dotted line) to these two disconnected arcs via Eq.(\ref{BCS-like-dispersion}) provides the value $\tilde{\Delta}(r_{\mathrm{max}})/E_{F} = 0.77$, in reasonable agreement with the value determined for the whole trap.
Figure~\ref{figS3} shows also the dispersion (dashed line) obtained by multiplying $A(k,\omega)$ by $f(\omega)$, in such a way that the low-$\omega$ peak can be smoothly followed even in the $k$-window that had to be excluded before.
This procedure does not appreciably alter the values obtained by the BCS-like fit.

This conclusion is consistent with the fact that the lack of a spectral depression in $A(k,\omega)$ in a \emph{limited\/} range of $k$ does not necessarily lead to disappearance of the pseudogap in integrated quantities, like the single-particle density of states (DOS) \cite{S-Ohashi-2009}, where a spectral depression survives at much higher temperature than in
 $A(k\approx k_{F},\omega)$ [cf. the inset of Fig.4(b) of the main paper, where a calculation of the DOS is explicitly reported].

The presence of the factor $f(\omega)$ in Eq.(\ref{theoretical-ECC-curves}) obviously affects more the large-$\omega$ than the low-$\omega$ peak of the EDC curves.
The discrepancies that are evident in the large-$\omega$ peak from Fig.~1 of the main paper, when comparing experimental and theoretical EDC curves at $T_{c}$
for the couplings $0.45$ and $0.57$, can accordingly be attributed to the larger absolute values of $T_{c}$ at which the theoretical spectra are calculated \cite{S-PPPS-2004}, with respect to the experimental values of $T_{c}$ at which the data are taken.

In Fig.~\ref{figS4} we reproduce the experimental EDC (circles) from Fig.~1 of the main paper for the two couplings $0.45$ and $0.57$, and compare them with the theoretical calculations 
(full lines) in which the temperature in the Fermi function has been decreased to $0.7 T_{c}$ while the temperature in $A(k,\omega)$ is kept at $T_{c}$.
This procedure is consistent with the fact that in this coupling range the theoretical approach overestimates the absolute value of $T_{c}$ by about $30 \%$, while close to $T_{c}$ the spectral function depends essentially on the relative temperature $T/T_{c}$.
This procedure, albeit empirical, goes in the right direction of reducing the height of the high-$\omega$ peak of the EDC curves making it closer to the experimental value, while affecting only
slightly the low-$\omega$ part of the EDC curves.

The numerical difference between the theoretical and experimental values of $T_{c}$ for a homogeneous Fermi gas at unitarity is also evident from Fig.~4(c) 
of the main paper, although this difference is immaterial to the sake of the argument that was there raised.

\begin{figure}[t]
\begin{center}
\includegraphics[angle=0,width=7.0cm]{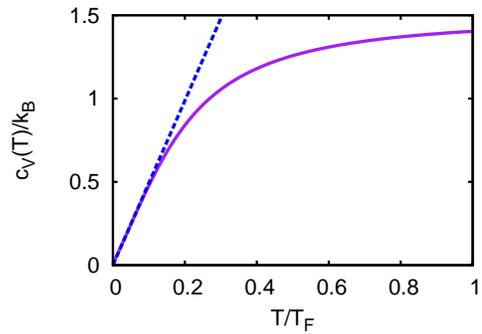}
\caption{Specific heat per particle of a non-interacting Fermi gas (full line) reported over an extended temperature range.
              The dashed line corresponds to the linear behavior that holds when $T/T_{F} \ll 1$.} 
\label{figS5}
\end{center}
\end{figure}

\begin{center}
{\bf Specific heat of a non-interacting Fermi gas} 
\end{center}
\vspace{-0.2cm}

In Fig.~4(c) of the main paper the behavior of the specific heat of a non-interacting Fermi gas was reported for comparison over the temperature interval $0.2 \lesssim T/T_{F} \lesssim 0.6$, 
which was relevant to the experimental data shown in the same figure.

It is interesting to show the same quantity over a more extended temperature range which reaches $T=0$.
This is done for $0 \le T/T_{F} \le 1$ in Fig.~\ref{figS5}, where the specific heat per particle of a non-interacting Fermi gas is reported vs $T/T_{F}$ (full line) and compared with its
\emph{linear} approximation $(k_{B} \pi^{2}/2) \, T/T_{F}$ (dashed line) that holds when $T/T_{F} \ll 1$.
Note that for $T/T_{F} = 0.2$ this linear approximation deviates from the full calculation already by about $20 \%$.

An analogous linear behavior is known to result for a Fermi liquid when $T/T_{F} \ll 1$, although with a different slope reflecting the renormalization of the mass \cite{S-PN-1996}.

\vspace{-0.3cm}



\begin{thebibliography}{99}
\vspace{-0.3cm}

\bibitem{Levin-2009-review} See, Q. Chen, Y. He, C. -C. Chien,  and K. Levin, Rep. Prog. Phys. {\bf 72}, 122501 (2009), and references therein.

 \bibitem{Randeria-1998} M. Randeria, in Proc. of the Intern. School of Physics ``Enrico Fermi'' Course CXXXVI on \emph{Models and Phenomenology for Conventional and High-temperature Superconductivity}, G. Iadonisi, J. R. Schrieffer, and M. L. Chiafalo, Eds. (IOS Press, Amsterdam, 1998), p. 53.

\bibitem{PPSC-2002} A. Perali, P. Pieri, G. C. Strinati, and C. Castellani, Phys. Rev. B {\bf 66}, 024510 (2002).

\bibitem{Salomon-2010} S. Nascimb\`{e}ne, N. Navon, K. J. Jiang, F. Chevy, and C. Salomon, Nature {\bf 463}, 1057 (2010).

\bibitem{Jila-Cam-2010-I} J. P. Gaebler, J. T. Stewart, T. E. Drake, D. S. Jin, A. Perali, P. Pieri, and G. C. Strinati, Nature Phys. {\bf 6}, 569 (2010).

\bibitem{Jin-2008} J. T. Stewart, J. P. Gaebler, and D. S. Jin, Nature {\bf 454}, 744 (2008).

 \bibitem{BCS-1957} J. Bardeen, L. N. Cooper, and J. R. Schrieffer, Phys. Rev. {\bf 108}, 1175 (1957). 

\bibitem{Luttinger-1960} J. M. Luttinger, Phys. Rev. {\bf 119}, 1153 (1960).  

\bibitem{SOM} For more details, see the ``supplemental material''.
   
\bibitem{Schneider-2010} W. Schneider and M. Randeria [Phys. Rev. {\bf 81}, 021601 (2010)] pointed out that the universal behavior 
                                          of a Fermi gas with a contact interaction \cite{Tan-2008} yields a weak negatively dispersing spectral feature 
                                          at $k\gg k_F$ even for a repulsive Fermi gas. In that case, however, this secondary peak cannot be traced down to $k\simeq k_F$. 

\bibitem{Tan-2008} S. Tan, Ann. Phys. {\bf 323}, 2971 (2008).

\bibitem{Levin-2009} Q. Chen and K. Levin, Phys. Rev. Lett. {\bf 102}, 190402 (2009).
                                                                
\bibitem{Bulgac-2009} P. Magierski, G. Wlazlowski, A. Bulgac, and J. E. Drut, Phys. Rev. Lett. {\bf 103}, 210403 (2009).                          

\bibitem{Ohashi-2009} S. Tsuchiya, R. Watanabe, and Y. Ohashi, Phys. Rev. A {\bf 80}, 033613 (2009); \emph{ibid.}  {\bf 82}, 033629 (2010).

 \bibitem{Nozieres-1964} P. Nozi\`eres, \emph{Theory of interacting Fermi systems} (Reading, MA, 1964). 

\bibitem{Zwerger-2009} R. Haussmann, M. Punk, W. Zwerger, Phys. Rev. A {\bf 80}, 063612 (2009).

\bibitem{Carlson-2005} J. Carlson and S. Reddy, Phys. Rev. Lett. {\bf 95}, 060401 (2005).

\bibitem{Shen-1998} F. Ronning \emph{et al.\/}, Science {\bf 282}, 2067 (1998).

\bibitem{KM-2000} C. Kusko and R. S. Markiewicz, Phys. Rev. Lett. {\bf 84}, 963 (2000).    

\bibitem{Shin-2010} Y-il Shin, Nature {\bf 463}, 1029 (2010).

\bibitem{Bulgac-2006} A. Bulgac, J. Drut, and P. Magierski, Phys. Rev. Lett. {\bf 96}, 90404 (2006).

\bibitem{Burovski-2006} E. Burovski, N. Prokofev, B. Svistunov, and M. Troyer, Phys. Rev. Lett. {\bf 96}, 160402 (2006).

\bibitem{Salomon-2010-II} S. Nascimb\`{e}ne, N. Navon, F. Chevy, and C. Salomon, 	arXiv:1006.4052v1.

\bibitem{SH-HTcSC-2009} H. -H. Wen \emph{et al.}, Phys. Rev. Lett. {\bf 103}, 067002 (2009).
                     
\end{thebibliography}

\begin{thebibliography}{99}
\vspace{-0.3cm}

\bibitem{S-PPSC-2002} A. Perali, P. Pieri, G. C. Strinati, and C. Castellani, Phys. Rev. B {\bf 66}, 024510 (2002).

\bibitem{S-PPS-2008} A. Perali, P. Pieri, and G. C. Strinati, Phys. Rev. Lett. {\bf 100}, 010402 (2008).

\bibitem{S-PPS-2009} P. Pieri, A. Perali, and G. C. Strinati, Nature Phys. {\bf 5}, 736 (2009).

\bibitem{S-Jin-2008} J. T. Stewart, J. P. Gaebler, and D. S. Jin, Nature {\bf 454}, 744 (2008).

\bibitem{S-Jila-Cam-2010-I} J. P. Gaebler, J. T. Stewart, T. E. Drake, D. S. Jin, A. Perali, P. Pieri, and G. C. Strinati, Nature Phys. {\bf 6}, 569 (2010).

\bibitem {S-LP-Stat_Phys} E. M. Lifshitz and L. P. Pitaevskii, \emph{Statistical Physics: Theory of the Condensed State} (Butterworth-Heinemann, 
                                       Oxford, 1980), Section 14.

\bibitem{S-PPPS-2004} A. Perali, P. Pieri, L. Pisani, and G. C. Strinati, Phys. Rev. Lett. {\bf 92}, 220404 (2004). 

\bibitem{S-Ohashi-2009} S. Tsuchiya, R. Watanabe, and Y. Ohashi, Phys. Rev. A {\bf 80}, 033613 (2009).

\bibitem{S-PN-1996} See, e.g., D. Pines and P. Nozi\`{e}res, \emph{The Theory of Quantum Liquids: Normal Fermi Liquids} (Addison-Wesley, Reading, 1996), Section 1.3.
                                
\end{thebibliography}
\end{document}